\documentclass[letterpaper]{article}

\usepackage{natbib,alifeconf}  
\usepackage{amsmath}
\usepackage{bbm}
\newcommand{\norm}[1]{\left\Vert #1\right\Vert}
\usepackage{url}

\usepackage{subcaption}
\title{Generating urban morphologies at large scales}
\author{Juste Raimbault$^{1,2,3,*}$ \and Julien Perret$^{1,4}$ \\
\mbox{}\\
$^1$UPS CNRS 3611 ISC-PIF\\
$^2$CASA, UCL\\
$^3$UMR CNRS 8504 G{\'e}ographie-cit{\'e}s\\
$^4$Univ. Paris-Est, LaSTIG STRUDEL, IGN, ENSG\\
\medskip
* juste.raimbault@polytechnique.edu} 

\begin{document}
\maketitle

\begin{abstract}
  At large scales, typologies of urban form and corresponding generating processes remain an open question with important implications regarding urban planning policies and sustainability.
  We propose in this paper to generate urban configurations at large scales, typically of districts, with morphogenesis models, and compare these to real configurations according to morphological indicators.
  Real values are computed on a large sample of districts taken in European urban areas.
  We calibrate each model and show their complementarity to approach the variety of real urban configurations, paving the way to multi-model approaches of urban morphogenesis.
\end{abstract}

\section{Introduction}

The study of forms of the built environment, and more precisely of the urban environment, has been the subject of different disciplines such as architecture, urban planning, or geography, with different approaches corresponding to various scales and processes \citep{moudon1997urban,gauthier2006mapping,kropf2009aspects}.
Establishing typologies of urban morphologies, and understanding their link with underlying urban growth processes, is nowadays a crucial issue for sustainability as a large majority of the world population live in cities and energy consumption is closely related to urban form through e.g. mobility patterns and automobile dependance \citep{newman2000sustainable}.

Although there is neither a unified definition of urban form, nor unified generative models and quantitative indicators to measure it, several approaches are close to the spirit of artificial life and generative social science~\citep{bonabeau1997classical,epstein1999agent}.
Procedural modeling \citep{watson2008procedural} aims at generating realistic cities, but is mostly focused on the visual impression given and does not consider realistic generative processes.
It is furthermore developed largely at larger scales than the one of the district \citep{Parish:2001:PMC:383259.383292}. \cite{merrell2010computer} generates in that context plans for interior of buildings, whereas \cite{cruz2017generation} uses a cellular automaton model for building morphogenesis.
Approaches linked to urban planning have focused on the spatial distribution of land-use, at multiple resolutions \citep{liu2017future}, and proposed cellular automata models for urban sprawl, generally at the scale of the metropolitan area \citep{herold2003spatiotemporal}. For example, \cite{horner2007multi} proposes a link between urban form based on land-use and commuting.

Urban form can furthermore be characterized considering different components of the urban system, such as building themselves as in several examples given before, but also for example transportation networks such as road networks \citep{ye2014quantitative}. To what extent these layers are complementary remains an open question, despite a few investigations coupling the two such as \cite{raimbault2018urban} suggesting indeed complementary dimensions. Regarding the geometrical properties of building layouts at the scale of a district, that we denote to simplify as urban form at a large scale, systematic characterizations and generative models remains rather rare. \cite{achibet2014model} for example describes a model of co-evolution of building layout and road network.

This paper proposes a first step towards a systematic understanding of generative models of the urban form, at a large scale.
The approach taken here is similar to the one taken by \cite{raimbault2018calibration}, which computes urban form indicators at a mesoscopic scale (metropolitan area) and calibrates a reaction-diffusion morphogenesis model.
We consider real urban configurations at the scale of the district (fixed spatial window of 500m), compute their morphological characteristics, and use these measures to calibrate different generative models of urban layouts at the same scale.
Our contribution is twofold: (i) we synthesize a set of indicators relevant at this scale, and compute them on a large sample of real urban configurations in European urban areas; (ii) we provide three different generative models complementary in the type of processes taken into account, and calibrate these models on the real morphological measures. We show therein the complementarity of the different processes to produce the variety of real urban forms considered.
This is to the best of our knowledge the first time several generative models at this scale are systematically compared on a large number of real configurations through quantitative measures.

The rest of this article is structured as follows.
First, we present the methods used in our work, including the measures allowing the comparison of urban forms, the proposed generative models and the method used to retrieve real urban configurations.
The results of the proposed approach are then explained together with the tools used in the calibration of the models.
Finally, the results are discussed.

\section{Methods} \label{sec:methods}

The approach taken requires both a robust way to quantify urban forms, through indicators that can be understood as features in the sense of machine learning, and generative models.

\subsection{Quantifying urban forms}

The quantification of urban form is in itself covered by a vast literature.
Recent work have proposed to apply deep learning techniques directly on vector data, such as \cite{2017arXiv170902939M} does, for a worldwide classification of road networks. Such an approach avoids the question of isolating relevant features.
However, as we aim at calibrating generative models, our quantification will make more sense with interpretable measures.
\cite{boeing2018measuring} proposes an extensive review of existing measures from a large extent of disciplines, their implications for planning and design, and the relation with urban complexity. \cite{webster1995urban} uses image processing techniques such as contrast or Fourier analysis, to extract synthetic descriptions of urban areas from satellite imaging. \cite{fumega2014identification} provide a typology of cities in relation with energy consumption in the perspective of climate change. \cite{rode2014cities} relate indicators of urban form with residential heat-energy demand. Other complexity-related approaches such as fractal dimensions have been introduced as for example by \cite{batty1987fractal}.

In practice, we use a variety of indicators capturing different aspects, each being detailed below. We consider the local urban space as a square grid of width $\sqrt{N}$ with cells $1 \leq i \leq N$, and an urban configuration is a binary function $s_i \in \{0;1\}$ on these cells. For the computation of indicators, we consider underlying complementary networks, the building network $B$ defined as nodes in centroids of occupied cells and links between two occupied direct neighbor cells (one cell unit of distance between centroids), and the free space network $\bar{B}$ defined similarly on empty cells. This raster representation is convenient as compatible with the various types of indicators and generators as described below. We will consider $\sqrt{N}=50$ in the following, and real windows of width $500m$.

\subsubsection{Basic indicators}
Simple descriptive indicators considered are (i) the total building density $A = \frac{1}{N}\cdot \sum_i s_i$; (ii) the number of buildings given by the number of connected components of $B$; (iii) the average building area, i.e. the average size of $B$ connected components; (iv) Moran index capturing spatial autocorrelation (see \cite{raimbault2018calibration} for its definition in a similar setting), with a simple inverse distance weight function; (v) average distance between non-empty points (which also captures a level of concentration).

\subsubsection{Network indicators}

We also use indicators computed with the underlying networks: the average detour computed in the free space network $\bar{B}$, computed by randomly sampling 50 pairs of points in a connected component of $\bar{B}$ and computing the ratio between the network distance and the euclidian distance $d_{\bar{B}}/d_E$. This measures captures in a way the sinuosity of streets from a mobility viewpoint. We also consider the average size of open connected areas as the average size of the connected components of $\bar{B}$.

\subsubsection{Mathematical morphology indicators}

Finally, indicators inspired from the field of mathematical morphology \citep{serra1983image} have already been applied to the quantification of urban form as for example by \cite{pesaresi2003recognizing}. Mostly used in image processing, these techniques proceed to the convolution of the image with a filter for example to simplify some morphological detail, what can be interpreted as a kind of spatial smoothing. We use here an indicator based on erosion with a filter of smallest size, which here removes points which 4 closest neighbors are not also occupied. We consider the total number of steps to fully erode the image, which is linked to building size. Similarly, using the operation of dilation, which in the contrary occupies points with at least one occupied neighbor, we consider the total number of dilation steps to fully fill the grid. This captures the size of open spaces. We do not consider indicators linked to opening and closing operations, as these would require more complex filters and for example their behavior as a function of kernel size.

Combining these morphological indicators, we have a total number of 9 indicators that can be computed on any binary grid, and that we will use in the following to compare real grids with generated grids of the same size.


\subsection{Generative models}

We detail now the generative model introduced. A null model is also considered, to ensure the relevance of the morphological measures and the adjustments on these. It consists in a random grid generator, where each cell is occupied if a random uniform number between 0 and 1 is below a density parameter $d_R$.

\begin{figure}
    \centering
    \includegraphics[width=\linewidth]{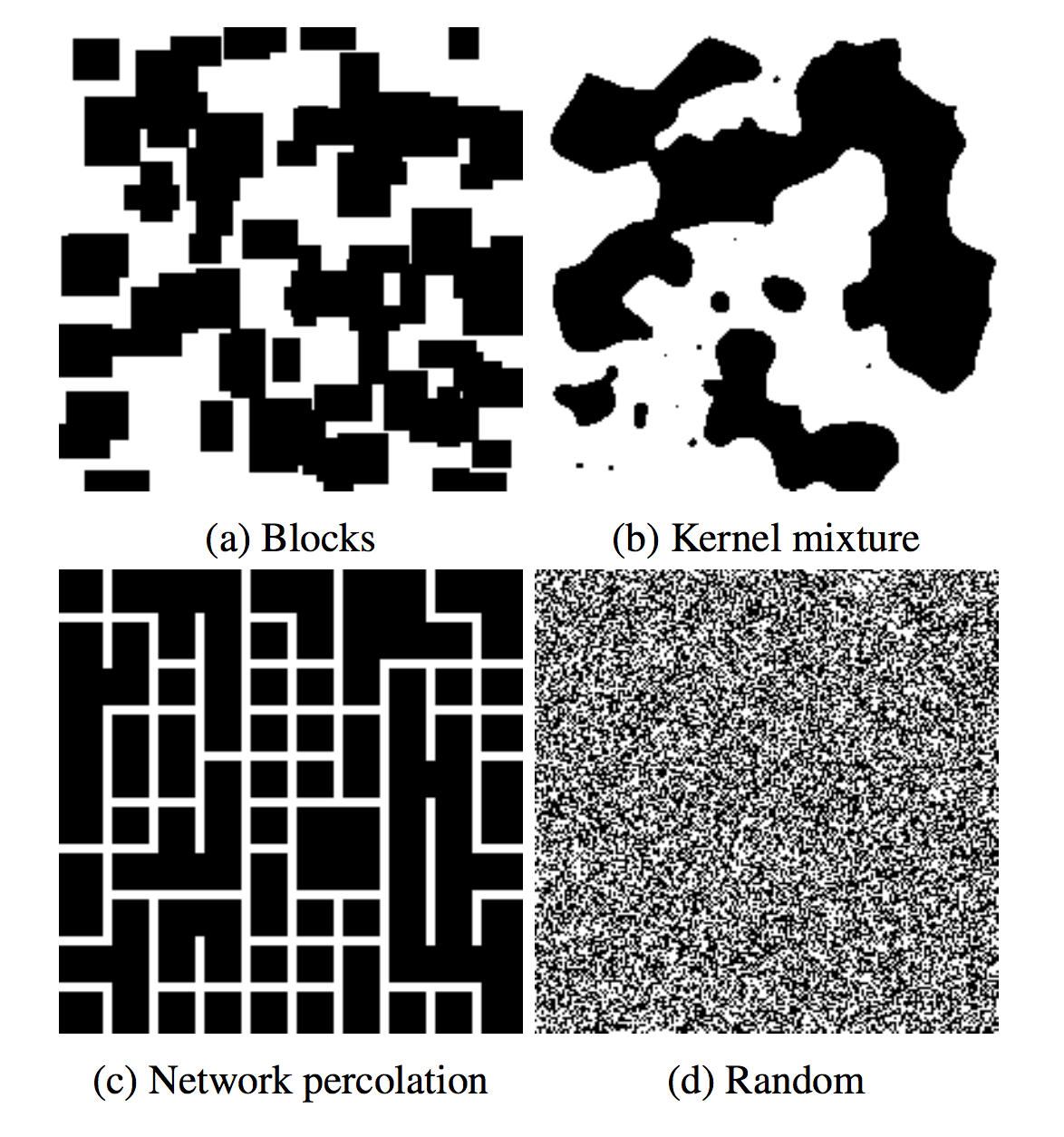}
    \caption{Examples of patterns produced by the synthetic generators.}
    \label{fig:generators}
\end{figure}

\subsubsection{Blocks generator}

The most simple ``realistic'' generator is similar to procedural modeling or marked point processes and distributes building blocks into the space (see Fig.~\ref{fig:blocks}). Given a number $N_B$ of blocks, random positions are drawn and block of a random height and width (with minimal value $m_B$ and maximal value $M_B$ as parameters) are placed at these.

\subsubsection{Kernel mixture generator}

Kernel mixture are a classical way to represent the spatial distribution of population density in an urban area \citep{anas1998urban} (see Fig.~\ref{fig:mixture}). They remain relevant at our scale, as they can be interpreted as a superposition of ``density hotspots'', as can the planning or the self-organization of a district can be. Given a number of centers $N_K$, $\vec{x}_{1\leq j \leq N_K}$ random position are drawn in the grid, at which kernels are applied, such that $s_i = \mathbbm{1}_{d_i \geq \theta_K}$ where the density $d_i$ for the point at position $\vec{y}_i$ is given by
\begin{equation}
    d_i = \frac{1}{N_K}\cdot \sum_j \exp{\left(-\norm{\vec{x}_j - \vec{y}_i}/d_K\right)} 
\end{equation}
where $d_K$ is a range parameter giving the extent of kernels.

\subsubsection{Network percolation model}

The last generator we used is based on network percolation, in the spirit of capturing the constraints imposed by flows traversing a given urban area (see Fig.~\ref{fig:percolation}). While the two previous generator were based on building processes, this one relies on streets, and thus on processes linked to transportation. The idea is to link a fixed number $N_P$ of border points, which can be understood as entrances/exits of the area. Starting with a grid network without links and nodes at a regular spatial sampling (fixed with a step of 5 units in our case), an iterative procedure (i) draws a random number and adds a random link at an empty potential link if it is smaller than a parameter called the percolation probability $p_P$; (ii) computes the largest connected component of the network and the number of nodes of this component on the boundary of the world; (iii) stops if this number is equal to the parameter $N_P$. Cells not covered by the resulting giant component are then occupied, at the exception of cells within a neighborhood $L_P$ of a link of the giant component. This way, this component can be understood as a circulating area linking $N_P$ entrances and exits, with a constraint on width through $L_P$.

Note that our generators will be ``fairly compared'' in terms of calibration, as they have the same number of parameters (although we do not introduce any information criteria that would yield the same penalization for overfitting).
We show in Fig.~\ref{fig:generators} visual representations of some outputs of each generator, including a random generator (see Fig.~\ref{fig:random}).

\section{Results} \label{sec:results}

Simulation results and real measures are available on the dataverse repository at \url{https://doi.org/10.7910/DVN/LGK0US}. Source code is available on the git repository of the project at \url{https://github.com/openmole/spatialdata}.
The model and indicators were coded in scala langage for performance purposes. This furthermore allows a seamless integration into the OpenMOLE workflow engine for model exploration \citep{reuillon2013openmole}, which provides methods for numerical experiments (in our case sampling methods) and transparent access to high performance computation environments (model simulation were run on the European Grid Infrastructure for an equivalent of one year and one month CPU time).

\begin{figure}
    \centering
    \includegraphics[width=\linewidth]{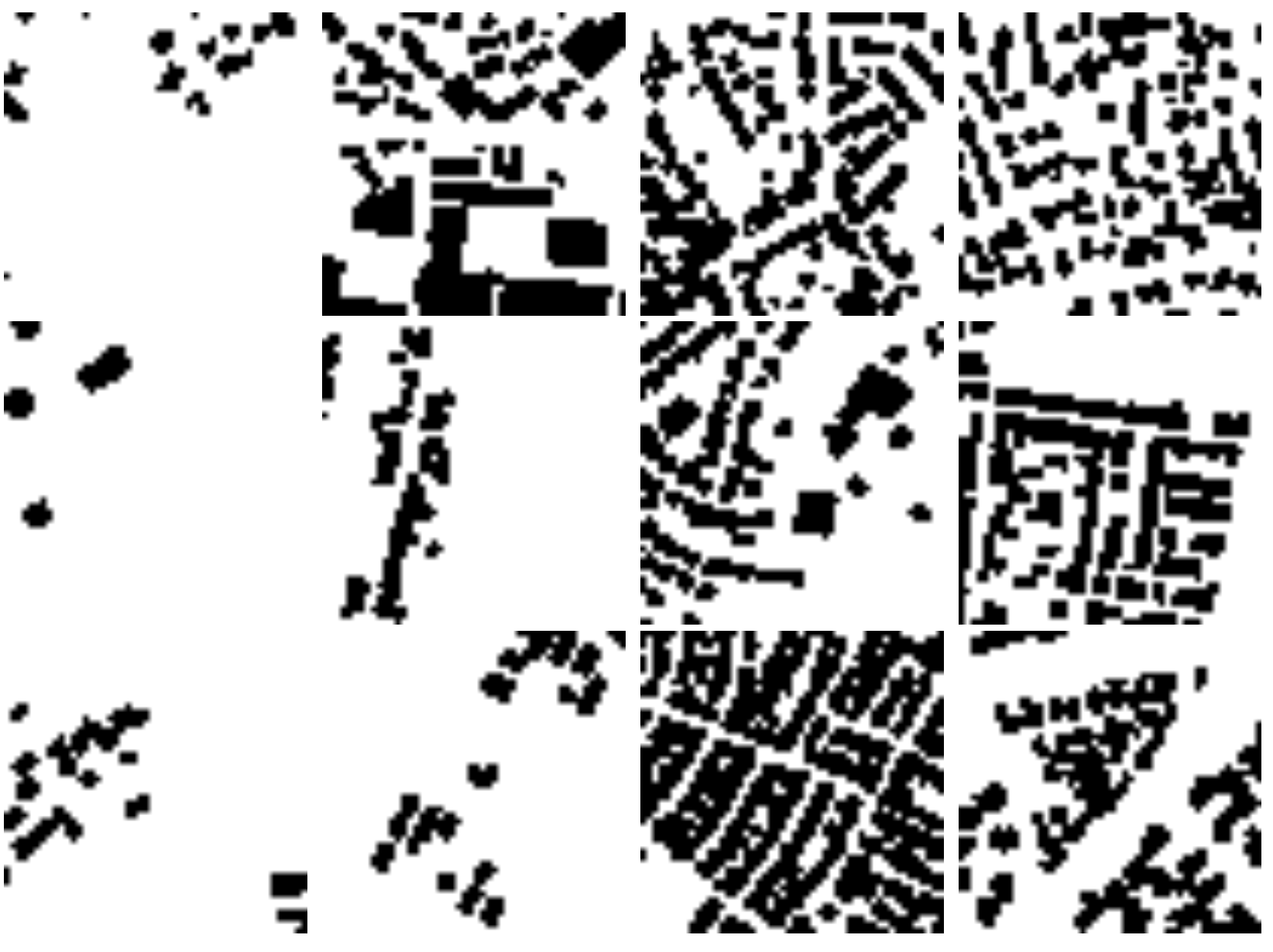}
        \caption{Samples extracted from OpenStreetMap.}
    \label{fig:osm}
\end{figure}

\subsection{Real measures}

\begin{figure*}
\vspace{-1cm}
  \centering
  \includegraphics[width=0.57\linewidth]{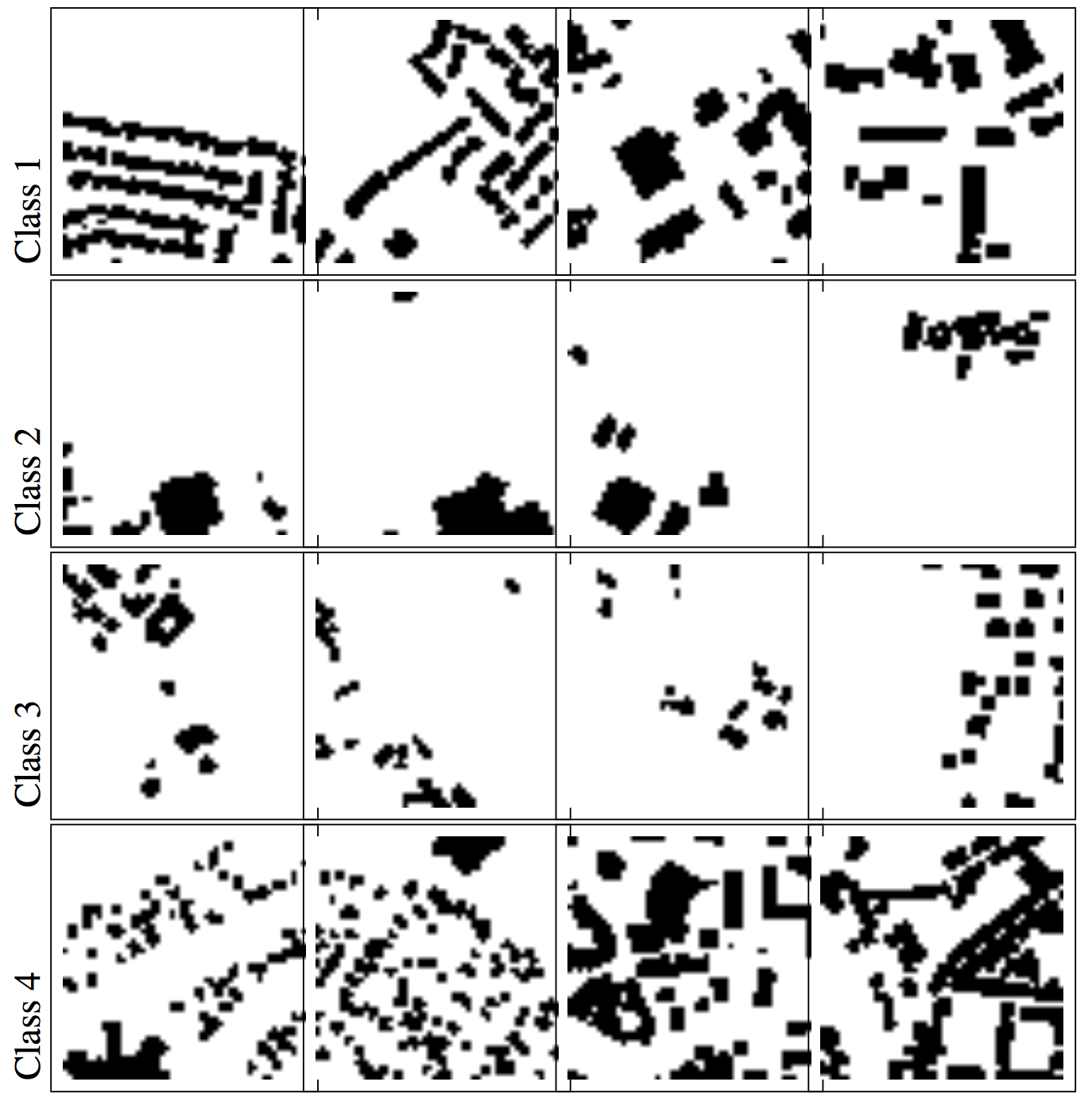}
    \caption{Typology applied to OpenStreetMap samples. The first class correspond mostly to a high density of linear buildings, recalling modern urbanism projects; the second disparate large buildings that can corresponds to industrials buildings in the outskirt of cities; the third disparate small buildings closer to periurban settlements; the fourth are denser and more complicated patterns evoking historical urban centers.}
  \label{fig:osm_types}
\end{figure*}

We compute the morphological indicators given above on a large sample on real urban areas. For practical computational reasons, we restrain our geographical area of study to European functional urban areas as provided by \cite{bretagnolle2019following}. We expect to already have a good representativity, although not universal, of existing urban forms with this sample, as it is known that European cities already have a significant morphological diversity \citep{le2015forme}. We collect building layouts from OpenStreetMap (illustrated in Figure~\ref{fig:osm}), as this source has been shown to have a good quality especially in Europe \citep{mooney2010towards}. Using the \texttt{osmosis} tool, buildings are filtered from the openstreetmap raw dump for Europe (downloaded from \url{http://download.geofabrik.de/} in March 2019) and inserted into a Postgis database, which can then be efficiently queried for a specific bounding box. Indeed, although the library developed provides a direct access to the OpenStreetMap API, query limitations do not allow such a systematic sampling.
We sample $N=72,000$ points into polygons corresponding to urban areas, first by selecting the area with a uniform selection weighted by population of areas, then by drawing uniform spatial coordinates within the polygon with a polygon sampling heuristic. 


After removing the empty areas and areas with a too low (lower than 0.05) or a too high (higher than 0.8) density, we end with $17,612$ real points on which the morphological measures are computed. The effective dimension is relatively low, echoing literature on urban form at other scales, as the first principal component on normalized indicators captures 70.3\% of variance, the second a cumulated proportion of 85.9\% and the third 92.8\%. The order of magnitude are similar to the ones found by \cite{Schwarz201029} for example. The first component captures low density (coefficient -0.43 for density) but clustered configurations (-0.35 for average distance), confirmed by the positive influence of dilation steps (0.44). On the contrary, the second component captures dispersed configurations (negative Moran and positive average distance) with large blocks (negative dilation steps).



We obtain a broad variety of forms, measures such as the Moran index varying between 0.02 (small disparate settlements) to 0.93 (one huge block), with a median at 0.10 (several medium size buildings). Similarly, the number of dilation steps varies from 3 (narrow streets only) to 80 (mostly open spaces) with a median at 26. The mean density is 0.21, what means that around 21\% of the soil is covered with building in average in the space we sampled, confirming that the most of urban areas are not dense contrary to the highly dense centers which are a minority.

To obtain typical representative points, we proceed to an unsupervised clustering on the two first principal components of these points. Using a k-means algorithm (5000 stochastic repetitions), varying the number of clusters shows an endogenous transition in the within-cluster variance proportion, suggesting to take $k=4$. Examples within the classes are shown and commented in Fig.~\ref{fig:osm_types}. The centroids will be used as typical objectives for model calibration.


\begin{figure*}
\vspace{-1.8cm}
    \centering
    \includegraphics[width=\linewidth]{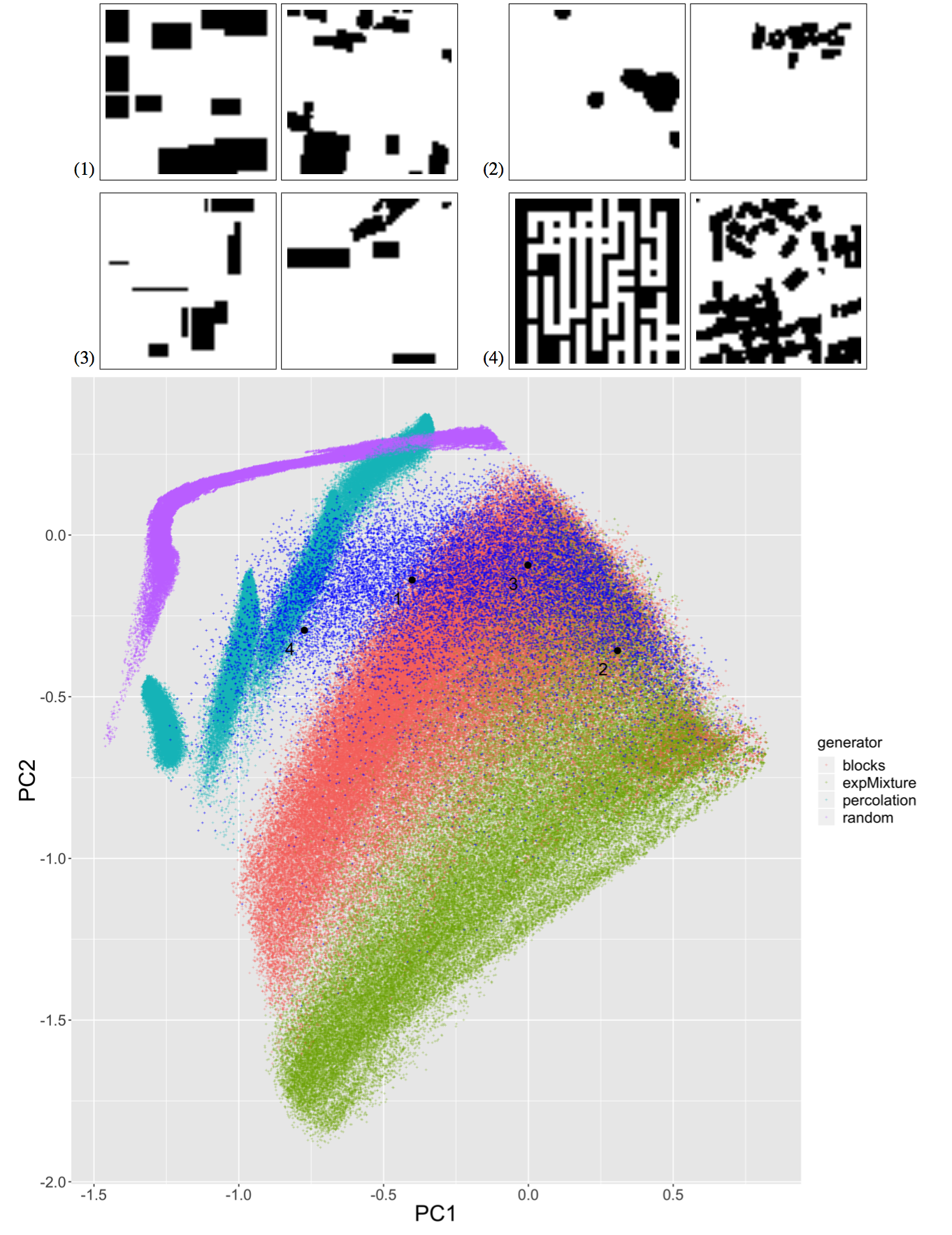}
    \caption{Comparison of real morphologies with patterns produced by the synthetic generators. All points are projected into the first two components computed on the real measures only. Points in dark blue correspond to real configurations. The type of generator is given by the color in legend. Dark circle correspond to the centroids of clusters on which the calibration is done, their number giving the corresponding configurations shown above. For each, we successively show the best synthetic configuration ((1) block generator with $N_B=12,m_B=5,M_B=12$, (2) exponential mixture generator with $N_K=6,d_K=3.8,\theta_K=0.42$, (3) block generator with $N_B=11,m_B=1,M_B=14$, (4) percolation generator with $p_P=0.35,N_P=5,L_P=3.7$), and the real configuration from OSM beside each.}
    \label{fig:lhs}
\end{figure*}

We can also consider the distribution of these measures within sampled urban areas. Keeping the areas with more than 10 sample points, we obtain 219 areas, for which we can compute the proportion of points within each morphological cluster. An Herfindhal diversity index on these proportions $p_k \in \left[0,1\right]$ computed as $h = 1 - \sum_k p_k^2$ ranges between 0.31 and 0.75 with an average of 0.63, suggesting very different profiles of urban areas. A chi-squared test of 10 levels of this diversity with the country is not significant (p=0.7), but the diversity index negatively correlates ($\rho = -0.11$, Fisher 95\% confidence interval $\left[-0.24,0.02\right]$) with longitude, meaning that Western cities are more diverse than Eastern cities, and more slightly with latitude ($\rho = 0.08 \left[-0.04,0.21\right]$).

\subsection{Model simulation and calibration}

A simulation experiment provides an insight into the patterns produced by the different generators in the morphological space. We sample the parameter space using a Latin Hypercube Sampling, with 10000 points for each generator respectively, and with 100 stochastic repetitions for each parameter point. This sampling is achieved with scripting the models into the OpenMOLE platform~\citep{reuillon2013openmole}.

Regarding the stochastic variability of generators, we compute for each indicator and each parameter point the sharpe ratios on repetitions, defined as the ratio between the estimated average and the estimated standard deviation. The indicators with the lowest values (high values indicate a low influence of stochastic fluctuations in comparison to variations due to parameters) are Moran index with a minimum of 0.22 and a median of 4.8, and the average detour with a minimum of 0.7 and a median of 5.2 (what could have been expected for this one as it is stochastically estimated). All other indicators have minimal sharpe ratios above 1.5 and medians above 5.4, meaning that models are overall not much sensitive to stochastic fluctuations. This confirms that considering single realisations as representing one parameter set remains reasonable.

We turn now to the comparison of generated configurations with real configurations. We work in the projected two dimensional space of the two first principal components of real points described above, in order to capture the maximum of variability in the real point cloud rather than in the simulated one. Note that working in the full indicator space makes no sense given the effective dimensions obtained (the simulated point cloud captures 93\% of variance at its third principal component, which is just a bit more than the real point cloud).

The point cloud of simulated and real points is shown in Fig.~\ref{fig:lhs}. We do not plot ensemble averages but all simulated points, as discussed above regarding the low influence of stochasticity. First of all, we observe that the null model consisting in random grids is far from all other points (except a tiny fraction of the percolation generator in turquoise) and in a way describes a boundary in the projected indicator space. This control confirms the relevance of projected indicators and of their comparison. 
Then, as expected since generators were conceived to capture different generative processes of the urban form, the point clouds of each generator are rather disjoint in the morphological space. The percolation generator produces separate clouds which correspond to different value of the link width $L_P$ parameter, and these are disjoint from the two other generators. The exponential mixture (green) and block (red) generators do overlap in a central area, but also have their own morphological ``exclusion zone'', where the forms can not be generated by other generators considered here.

When looking at the real point cloud, we see that most of it is covered by some generated points, and that generators are complementary to approach all covered points. This is an important result in line with the targeted complementarity of generative processes, and advocates for multi-modeling in urban morphogenesis. Interestingly, there is an area not covered, corresponding to the transition between the percolation generator (narrow streets) and the block generator.

For each centroid of the clusters in the real point cloud described above, that can be considered as a typical calibration objective, we provide example of the closest real configuration and the closest simulated one. Visually, forms are rather satisfying, at the exception of the percolation generator fitting a complicated urban center. Indeed, this centroid (number 4) is at the boundary of the percolation point cloud, and the real point cloud is more difficultly captured in this area compared to the block and mixture generators.


\begin{table*}[h]
\center{
\begin{tabular}{|c|c|c|c|c|}\hline
           & Random & Blocks & Exp. Mixture & Percolation\\ \hline\hline
Centroid 1 & $0.424 \pm 0.011$ & $0.106 \pm 0.063$ & $0.303\pm 0.101$ & $0.325\pm 0.019$ \\
Centroid 2 &$0.809\pm 0.022$&$0.164\pm 0.099$&$0.184\pm 0.141$&$0.947\pm 0.019$\\
Centroid 3 &$0.428\pm 0.019$&$0.095\pm 0.054$&$0.109\pm 0.064$&$0.541\pm 0.019$\\
Centroid 4 &$0.515 \pm 0.005$&$0.311\pm 0.077$&$0.589\pm 0.149$ &$0.083\pm 0.025$\\\hline
\end{tabular}
}
\vskip 0.25cm
\caption{Aggregated distance in the morphological space, for each generator and each calibration objective (real clusters centroids). Euclidian distance in the projected space are aggregated in average on stochastic repetitions, and the minimal average value is reported with its standard deviation.\label{tab:averagedists}}
\end{table*}

To quantify the level of calibration of each generator regarding each centroid, and in average regarding stochastic repetitions, we provide in Table~\ref{tab:averagedists} the aggregated minimal values of distances, for each generator and each calibration objective, with their standard deviations. This mainly confirms the previous results, with however interesting variations: (i) for the second centroid, the exponential mixture is in average no longer the best, and furthermore has a higher variability; (ii) centroid three and four are the easiest to reach, despite the latest being in the boundary; (iii) the percolation generator performs well on this point and has a very low variability.

This experiment has therefore shown the possibility to calibrate the generative models on morphological measures against real configurations, and furthermore unveils their complementarity to approach the diverse existing forms.

\section{Discussion} \label{sec:discussion}


Our approach provides a first step towards systematic modeling of generative processes of urban form at large scales. Some direct limitations could be tackled in a short term. Testing slightly different processes and heuristics in generator may be a way to cover the part of the real point cloud which is missed by our generators. As it seems correspond to complicated urban centers, it may be however complicated without more elaborated models. Also, we did not use minimization algorithms to calibrate the generators, and the and a further step would consist in checking the robustness of our result using such optimization heuristics (genetic algorithms for example), but also diversity algorithms such as pattern space exploration proposed by \cite{10.1371/journal.pone.0138212}, to ensure the effective feasible space of each generator.

This work can also be extended in several ways. First of all, we focused on the built environment but neglected transportation infrastructures, whereas spatial network morphogenesis models have been proposed for example by \cite{courtat2011mathematics} or \cite{raimbault2018multi} in a multi-modeling approach.
Taking into account multiple dimensions of the urban system is an important extension and hybrid models such as co-evolution models \citep{raimbault2014hybrid} should be investigated.

Furthermore, we tested the complementarity of generators only in a static way. Adaptive and dynamic generators, combining processes of different nature within the same model with an endogenous switching or combination, would be an important direction to better understand urban morphogenesis.
In the same context, the generators compared here had all the same number of parameters, but richer generators implying different numbers would require the use of information criterions to avoid overfitting, which, however, remains an unsolved issue for such generative simulation models \citep{piou2009proposing}.

Finally, as extensively reviewed above, the way to quantify urban form strongly depends on the scale considered. A more integrative understanding of it would require multi-scale approaches able to relate these different definition and measures within a single multi-scalar framework.

\section{Conclusion}

We have proposed here a new insight into the generative simulation of urban morphologies at large scales, namely the scale of the district considering the layout of buildings. After computing morphological measures on a large sample of real urban areas, we showed the complementarity of different generators capturing various aspects of urban morphogenesis processes. Despite not implying generative agents (developers, inhabitants, companies) and thus staying close to procedural modeling, this work however paves the way towards a more systematic understanding of generative processes of urban form at this scale.

\section{Acknowledgements}

Results obtained in this paper were computed on the vo.complex-system.eu virtual organization of the European Grid Infrastructure ( http://www.egi.eu ). We thank the European Grid Infrastructure and its supporting National Grid Initiatives (France-Grilles in particular) for providing the technical support and infrastructure.

\end{document}